\documentclass[aps,prl,preprint]{revtex4-1}

\usepackage{graphicx, times}
\usepackage{epstopdf}
\usepackage{bm}

\begin{document}

\title{Robustness of Optimal Synchronization in Real Networks}

\author{Bhargava Ravoori}
\affiliation{Institute for Research in Electronics and Applied Physics, University of Maryland, College Park, MD 20742}
\affiliation{Department of Physics, University of Maryland, College Park, MD 20742}

\author{Adam B. Cohen}
\affiliation{Institute for Research in Electronics and Applied Physics, University of Maryland, College Park, MD 20742}
\affiliation{Department of Physics, University of Maryland, College Park, MD 20742}

\author{Jie Sun}
\affiliation{Department of Physics and Astronomy, Northwestern University, Evanston, IL 60208}

\author{Adilson~E.~Motter}
\affiliation{Department of Physics and Astronomy, Northwestern University, Evanston, IL 60208}

\author{Thomas E. Murphy }
\affiliation{Institute for Research in Electronics and Applied Physics, University of Maryland, College Park, MD 20742}
\affiliation{Department of Electrical and Computer Engineering, University of Maryland,College Park, MD 20742}

\author{Rajarshi Roy}
\affiliation{Institute for Research in Electronics and Applied Physics, University of Maryland, College Park, MD 20742}
\affiliation{Department of Physics, University of Maryland, College Park, MD 20742}
\affiliation{Institute for Physical Science and Technology, University of Maryland, College Park, MD 20742}


\begin{abstract}
Experimental studies of synchronization properties on networks with controlled connection topology can provide powerful insights into the  physics of complex networks. Here, we report experimental results on the influence of connection topology on
synchronization in fiber-optic networks of chaotic optoelectronic oscillators. We find that the recently predicted non-monotonic,
cusp-like synchronization landscape manifests itself in the rate of convergence to the synchronous state. We also observe that networks with the same number of nodes, same number of links, and identical eigenvalues of the coupling matrix can exhibit fundamentally different approaches to synchronization. This previously unnoticed difference is determined by the degeneracy of associated eigenvectors in the presence of noise and mismatches encountered in real-world conditions.
\end{abstract}

\maketitle

Recent research~\cite{DorogovtsevRMP2008} has shown that network structure plays a significant role in cascading failures
\cite{WattsPNAS2002}, epidemics~\cite{PastorSatorrasPRL2001}, and recovery of lost network function~\cite{Motter_MSB2008}. Synchronization of coupled dynamical units is a widespread phenomenon that has served as an example \emph{par excellence} of this line of theoretical research \cite{ArenasPR2008}. For example, by modeling network synchronization in terms of diffusively-coupled identical oscillators, it has been shown that the stability of fully synchronous states is entirely determined by the eigenvalue spectrum of the coupling matrix~\cite{Pecora1998, Nish2006,*Nish2006PD}. A fundamental yet largely unexplored question concerns the robustness of such network-based predictions.

New insight into this question has been provided by a recent study on networks that optimize the synchronization range
\cite{Nish2010}. It is predicted that synchronization properties, such as the coupling cost at the synchronization threshold and range of coupling strength for stability, will exhibit a highly non-monotonic, cusp-like dependence on the number of nodes and
links of the network \cite{Nish2010}, contrary to the prevailing paradigm. The existence of such cusps indicate that small
perturbations of the network structure, which might be experimentally unavoidable, may lead to large changes in the network dynamics.

In this work, we experimentally demonstrate that the rate of convergence to synchronous states, a broadly significant synchronization property, follows the theoretically predicted non-monotonic trend. More important, we observe that networks with identical eigenvalue spectra (generally assumed to behave in similar fashion) can exhibit qualitatively different convergence properties.  We classify these networks into two groups, which we term {\it nonsensitive networks} and {\it sensitive networks}, respectively. This classification is based on the properties of the {\it eigenvectors} of the coupling matrix and the observation that networks with different eigenvector degeneracies will respond differently to perturbations typical of experimental conditions. In contrast to sensitive networks, nonsensitive networks are predicted and experimentally observed to be robust against these perturbations.  We identify observational noise and mismatch of coupling strengths as the main experimental factors underlying these different responses.

\begin{figure}[b]
\centering
\includegraphics{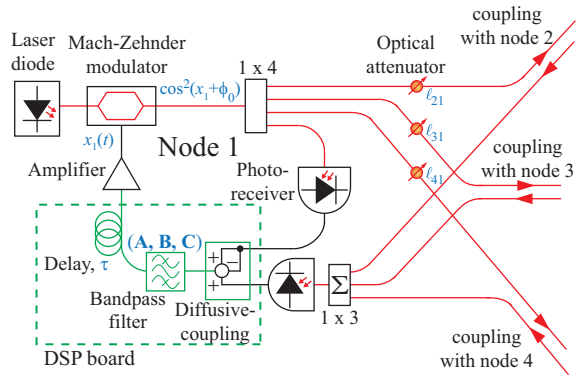}
\caption{Schematic of a single optoelectronic node. Each node is
coupled to the rest of the network (not shown) through fiber-optic
links. The couplings are enabled/disabled using optical
attenuators. } \label{fig1}
\end{figure}

Our experimental setup consists of a network of $N=4$ optoelectronic feedback loops. The feedback loops are similar in construction to those used by Argyris \emph{et al.} for chaotic communication \cite{Argyris2005}. Each feedback loop (Fig.\
\ref{fig1}) comprises a semiconductor laser which provides a steady optical power, a Mach-Zehnder electro-optic intensity
modulator, two photoreceivers, a digital signal processing (DSP) board which provides electronic filtering and time delay, and an amplifier. The optical output of each electro-optic modulator is proportional to $\cos^2(x_i+\phi_0)$, where $x_i$ is the
normalized electrical input voltage that characterizes each oscillator, and $\phi_0$ is the operating point of the modulator.
The signals $x_i(t)$ are recorded using a four-channel, 8-bit digital oscilloscope. The modulator output is split to act as the
feedback signal and as the coupling signal to the other nodes, with each coupling  link either enabled or disabled using optical
attenuators. In our experiments, all the couplings are set to have the same strength. At each node, the feedback and the coupling signals are processed using the DSP board. The parameters of each loop are set such that the oscillators exhibit high-dimensional chaos. The DSP board implements a 2-pole digital bandpass filter and a time delay on both the feedback and coupling signals. A diffusive-coupling scheme is implemented through the DSP board.
The equations that describe each node in the experimental network are derived in Ref.~\cite{Murphy2010}, and are given by:
\begin{eqnarray}
\frac{d\mathbf{u}_i(t)}{dt} &=& \mathbf{A} \mathbf{u}_i(t)-\mathbf{B} \beta \cos^2[x_i(t-\tau)+\phi_0], \label{eq1}  \\
x_i(t) &=& \mathbf{C} \Big( \mathbf{u}_i(t)- \frac{\epsilon}{d} \sum_{j}\ell_{ij}\mathbf{u}_j(t)\Big), \label{eq2}\\
\nonumber
\mbox{where }  \mathbf{A}&=&\left[ \begin{array}{cc}
-(\omega_1+\omega_2)& -\omega_2 \\
\omega_1& 0\\ \end{array} \right],~
\mathbf{B} = \left[ \begin{array}{c}
\omega_2 \\
  0\\ \end{array} \right],~
\mathbf{C}=\left[ 1 \mbox{ } 0 \right].
\end{eqnarray}
Here, $\mathbf{u}_i(t)$ is a $2\times1$ vector describing state of the filter at node $i$, and $x_i(t)$ is the observed variable. The oscillators are diffusively coupled through the network specified by the coupling matrix $\mathbf{L}=(\ell_{ij})$; the diagonal
element $\ell_{ii}\ge 0$ is the net incoming coupling strength to node $i$ and the off-diagonal element $\ell_{ij}$ is the negative
of the directional interaction strength from node \emph{j} to node \emph{i}.
Thus, if there is a link from $j$ to $i$, the influence of oscillator $j$ on oscillator $i$ is proportional to $[u_j(t)-u_i(t)]$.
Matrices $\mathbf{A}$, $\mathbf{B}$, and $\mathbf{C}$ represent the filter in state-space. The filter band is from
$\omega_1/2\pi=0.1$kHz to $\omega_2/2\pi=2.5$kHz. Regarding the other parameters, $\beta=3.6$ is a lumped effective feedback strength that combines the gain factors of various components, $\epsilon$ is a global coupling strength,
$d\equiv\mbox{Tr}(\mathbf{L})/N$ is a normalization factor defined by the average coupling  per node,  $\phi_0$ is a phase bias set to $\pi/4$, and $\tau=1.5$ms is the net feedback delay. Equations~(\ref{eq1}-\ref{eq2}) are a network generalization of the one- and two-oscillator systems considered in Refs.\ \cite{Murphy2010,Yanne2005,*Cohen2008}. This network model admits synchronous solutions $x_1(t)=x_2(t) \dots =x_N(t)$, whose experimental realization is the focus of this study.

\begin{figure}[b]
\centering
\includegraphics[width=3in]{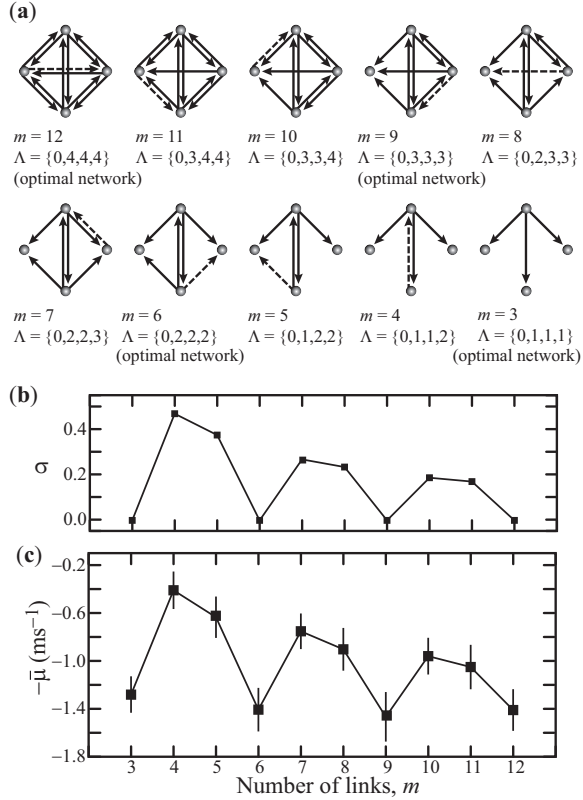}
\caption{Non-monotonic behavior of synchronization properties.
(a) A path from a fully connected network ($m=12$) to an optimal tree network ($m=3$). At each step, the link removed is indicated by a dashed line.
(b) The eigenvalue spread $\sigma$ for the networks in (a).
(c) Experimentally measured mean convergence rate to synchronization, $\bar{\mu}$,  and associated standard deviation
(bars) for the same networks. The largest rate of convergence is observed when the number of network links is a multiple of
$N-1=3$.
} \label{fig2}
\end{figure}

Consider a network of $N$ oscillators and  $m\equiv\mbox{Tr}(\mathbf{L})$ directed links, of which our experimental system is an example. Since all the rows of matrix $\mathbf{L}$ sum to 0, $\mathbf{L}$ has at least one null eigenvalue. The eigenvalue spectrum $\Lambda =\{0,\lambda_2,\lambda_3,\dots,\lambda_{N}\}$ of $\mathbf{L}$ determines whether the synchronous solutions for a given network configuration are stable \cite{Pecora1998,Nish2006,*Nish2006PD}.
Consider the eigenvalue spread~\cite{Nish2010},
\begin{equation}
\sigma^2 \equiv
\frac{1}{d^2(N-1)}\sum_{i=2}^{N}|\lambda_i-\overline{\lambda}|^2
\label{eq3}, \mbox{ where } \overline{\lambda} \equiv
\frac{\sum_{i=2}^{N}\lambda_i}{(N-1)},
\end{equation}
which measures the range of coupling strength $\epsilon$ for stable synchronization and hence the {\it synchronizability} for
general directed networks. Smaller eigenvalue spread implies higher synchronizability. Focusing on networks with the smallest
eigenvalue spread for given number of nodes and links, Ref.\ \cite{Nish2010} shows that the  eigenvalue spread has cusp-like
minima with $\sigma=0$ when $m=k(N-1)$  and  $\sigma>0$ for all networks with $k(N-1)<m<(k+1)(N-1)$, where $k=1,2,\dots,N$. The networks minimizing $\sigma$ for a given number of nodes and links are termed {\it optimal} if $\sigma=0$ and {\it suboptimal} if $\sigma>0$ (all the others are termed {\it nonoptimal}).  In Fig.\ \ref{fig2}(a), we show a sequence of 4-node optimal and suboptimal networks with decreasing number of links, which are considered in our experiment.  The eigenvalue spread $\sigma$ of these networks exhibit pronounced non-monotonicity as a function of the number of links [Fig.\ \ref{fig2}(b)].

In our experimental study, we first focus on the influence of network structure on synchronization properties. To this end we consider stable synchronous states, which correspond to configurations for which the synchronization error, defined as
\begin{equation}
\theta(t) \equiv \frac{1}{N(N-1)}\sum_{i,j}|x_i(t)-x_j(t)|, \label{eq4}
\end{equation}
ideally approaches zero. For real networks that synchronize, this error converges to a synchronization floor, $\theta_0$, determined by experimental mismatches and noise.  Figure \ref{fig2}(c) shows the experimentally measured rate of convergence to synchronization for the sequence of optimal and suboptimal networks shown in Fig.\ \ref{fig2}(a).
This rate of convergence is defined as the exponent $\mu$ of the exponential decay to the synchronization floor, $(\theta-\theta_{0})\sim \exp({-\mu}t)$,
and can in principle be determined from the  eigenvalues of the coupling matrix scaled as $\{\frac{\epsilon}{d}\lambda_i\}$ and the master stability function~\cite{Pecora1998}, which is defined by the node dynamics, form of the coupling, and synchronous state.
Before time $t=0$, the nodes are uncoupled and evolve independently. At time $t=0$, the couplings in the network are enabled by switching $\epsilon$ from $0$ to $0.7$, and the network converges to a synchronous solution.  In order to avoid problems with zero crossings, we perform a boxcar moving-average over a small time interval on $\theta(t)$ to form
$\langle\theta(t)\rangle$. We measure  $\mu$ by fitting the smoothened synchronization error $\langle\theta(t)\rangle$ to an
exponential over a fixed time interval from $0.5$ms to $2.0$ms. For each network, our statistics is based on performing this
measurement for $100$ independent realizations of the initial conditions. The results shown in Fig.\ \ref{fig2}(c) indicate only
small variability across different realizations. More important, contrary to what has been usually assumed, the measured mean
convergence rate $\bar{\mu}$  is found to change highly non-monotonically, with periodic peaks at the points where the number of links is a multiple of $(N-1)$ \cite{Nish2010}. The eigenvalue spread $\sigma$ is seen to be inversely related to the convergence rate to synchronization, i.e., the larger the spread, the slower the approach to synchronization. Results for larger
networks are included in supplementary information, Fig.~S1.

\begin{figure}[t]
\centering
\includegraphics{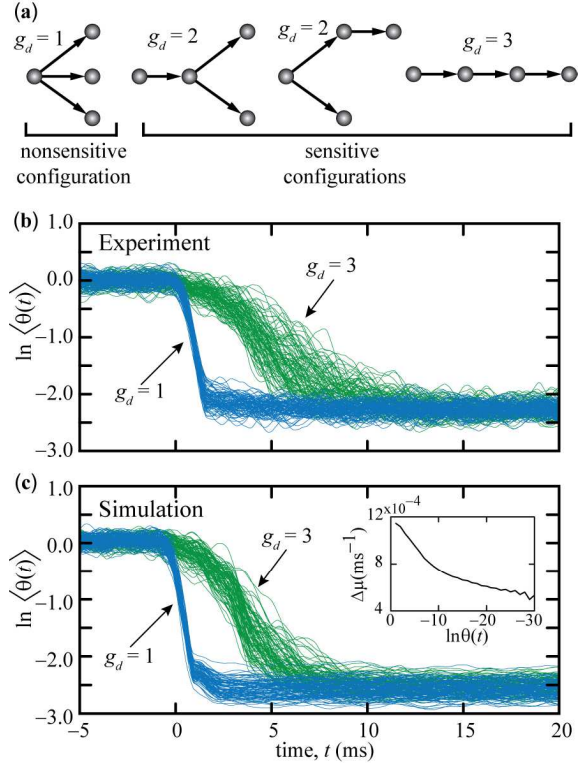}
\caption{Differentiating behavior between sensitive and nonsensitive networks.
(a) All optimal binary networks with $4$ nodes and $3$ links. Each network is labeled according to its geometric degeneracy, $g_d$.
(b) Experimentally measured $\langle\theta(t)\rangle$ for sensitive (green, $g_d=3$) and nonsensitive (blue, $g_d=1$) configurations, where the coupling is enabled at $t=0$. The individual curves represent measurements repeated for $100$ different initial conditions for each network.
(c) Numerical simulation of the same networks and conditions considered in (b).
Inset: difference $\Delta\mu$ between the decay exponents of the networks considered in (b) when simulated in the absence of mismatches, noises, and time delays, as a function of $\theta(t)$, regarded as a tunable initial synchronization error.
}\label{fig3}
\end{figure}

If experimental noises, delays and mismatches could be neglected, the synchronization properties would be entirely determined by the eigenvalues of the coupling matrix~\cite{Pecora1998,Nish2006,*Nish2006PD}. In particular, each network in the sequence of Fig.\ \ref{fig2}(a) is characterized by eigenvalues that minimize the spread $\sigma$.  The sequence of optimal and suboptimal networks considered in our experiments of Fig.\ \ref{fig2}(c) was generated by starting from a fully connected network and successively removing links while keeping the coupling matrix diagonalizable, so that the stability of the synchronous states could be analyzed within the standard master stability approach \cite{Pecora1998}. However, the coupling
matrices of directed networks are not necessarily diagonalizable. There are in fact many more optimal and suboptimal networks with the exact same eigenvalues of those considered in Fig.\ \ref{fig2}(a), but that are not diagonalizable because they have a number of independent eigenvectors smaller than $N$~\cite{Nish2006,*Nish2006PD}. For instance, out of four 4-node optimal networks with three links [Fig.\ \ref{fig3}(a)], one is diagonalizable and three are nondiagonalizable. (For the set of
all optimal and suboptimal binary 4-node networks, see supplementary information, Table~S1.)  Given that  $\sigma$
depends only on the eigenvalues, one might expect that experimental realizations of nondiagonalizable networks would
exhibit properties similar to those observed for the diagonalizable counterparts.

In Fig.\ \ref{fig3}(b), we experimentally compare the approach to synchrony of two networks,  a directed star and a directed linear chain, which have the  maximum and minimum number of independent eigenvectors, respectively. These two networks are optimal and have the same number of nodes and links and identical eigenvalues. We performed $100$ independent measurements of $\langle\theta(t)\rangle$ starting with different initial conditions for both networks. However, both the convergence to synchronization and the oscillations after synchronization are systematically different for these two networks. We refer to networks with nondiagonalizable coupling matrices as {\it sensitive networks}, since the experiments show that they are more susceptible to the influence of imperfections typical of realistic conditions. On the other hand, networks with diagonalizable coupling matrices are found to be fairly robust under the same conditions, and are referred to as {\it nonsensitive networks}. Mathematically, these two different types of networks can be categorized according to their {\it geometric degeneracy}, $g_d$, which is the largest number of repeated eigenvalues of the coupling matrix associated with the same (degenerate) eigenvector. In the star network,  each eigenvalue is associated with a linearly independent eigenvector, and hence $g_d=1$. In the case of the linear chain, all three nonzero eigenvalues are associated with the same eigenvector, and hence $g_d=3$. While we focus on optimal and suboptimal networks, where sensitive networks are expected to be more common because of their highly degenerate eigenvalue spectra, this classification also applies to nonoptimal networks in general.

Compared to the nonsensitive case, the sensitive networks exhibit slower convergence to synchronization and, across different realizations, larger variations around the average synchronization trajectory [Fig.\ \ref{fig3}(b)]. In particular, while the nonsensitive network has an exponential convergence to synchronization, the sensitive network has a non-exponential
convergence,  which is in agreement with the polynomial transient theoretically predicted for such networks~\cite{Nish2006,*Nish2006PD}. Moreover, the bundle of trajectories $\theta(t)$ is broader by a factor of ten for the sensitive
network over the nonsensitive network in the transient to synchronization. This difference, we hypothesize, is due to the
different response exhibited by these different types of networks to experimental perturbations, since in the absence of mismatches, noises, and delays, the asymptotic rate of convergence is expected to be the same. The latter is confirmed in the inset of Fig.\ \ref{fig3}(c), where we simulated Eqs.\ (\ref{eq1}-\ref{eq2}) under these conditions, thus recovering the prediction previously established for the idealized Pecora-Carroll model~\cite{Nish2010}. The essence of this phenomenon is that
convergence to zero is the same for both types of networks, but to a finite synchronization floor it is not.

\begin{figure}[t]
\centering
\includegraphics[width=3.4in]{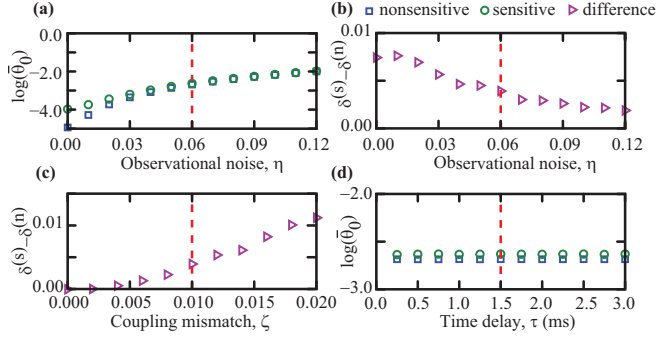}
\caption{Dependence of synchronization  properties on experimental parameters.
Setting observational noise $\eta=0.06$, coupling mismatch $\zeta=0.01$, and time delay $\tau=1.5$ms (dashed lines),
which approximate the values estimated in the experiment, we simulated the effect of varying one of these parameters at a time. Each data point is estimated from 1000 independent realizations for the networks considered in Fig.~\ref{fig3}(b,c).
(a) effect of $\eta$ on the ensemble mean synchronization floor, $\overline{\theta}_0$;
(b) effect of $\eta$ on the standard deviation of the floor, ${\delta}$;
(c) effect of $\zeta$ on ${\delta}$; and (d) the effect of $\tau$ on $\overline{\theta}_0$.
The superscript $n$ ($s$) denotes the nonsensitive (sensitive) network.
} \label{fig4}
\end{figure}

To test our hypothesis, we have simulated Eqs.\ (\ref{eq1}-\ref{eq2}) in the presence of observational noise and coupling mismatch. The coupling mismatch is taken to be independent perturbations to the nonzero off diagonal elements of $\ell_{ij}$ in Eq.\ (\ref{eq1}) drawn from a Gaussian distribution with zero mean and standard deviation $\zeta$. The observational noise is modeled as the difference between the actual $x_i(t)$ in the system, described by Eq.\ (\ref{eq2}), and the observed $x_i(t)$, drawn from a Gaussian distribution with zero mean and standard deviation $\eta$. Based on the dominant factors in the
experimental setup, we choose these values to be $\eta$ = 0.06 and $\zeta$ = 0.01, which are estimates for the rounding error in the recording of the data and coupling mismatch due to realistic imperfections in the network construction. As shown in Fig.\ \ref{fig3}(c) (and, for larger networks, in supplementary information, Fig.~S2), with this parameter choice our simulation of the system mimics the key features observed in the experiment to a remarkable degree.

The parameter dependence is further investigated in Fig.\ \ref{fig4}, where we simulate the dependence of the average
synchronization floor and the variation around it for sensitive and nonsensitive configurations, as a function of the noise,
$\eta$, mismatch, $\zeta$, and the feedback delay time, $\tau$. The floor itself is mainly determined by the observational noise.
The difference in the variations around the floor is mainly determined by the coupling mismatch. The time delay, on the other
hand, is found to have very limited influence on these properties.
As shown in supplementary information, Fig.~S3, similar results hold for larger networks.
Our simulations also show that oscillator mismatch and dynamical noise comparable to $\zeta$ and $\eta$ would lead to very large difference between the average synchronization floor of sensitive and nonsensitive networks; since this is not observed
experimentally we posit that these two factors  are likely to be extremely small in the experiment. Incidentally, this also
illustrates the distinct nature of the problem considered in this study compared to eigenvector-dependent synchronization in
externally forced systems~\cite{Kori2009} and in systems with oscillator mismatches~\cite{Restrepo2004,Sun2009}. On the other hand, while we classify our networks according to degeneracy of the eigenvectors, we note that nonsensitive networks can exhibit different levels of nonnormality, ranging from the extreme in which all eigenvectors are orthogonal to the case in which two or more of them are nearly parallel. Among the nonsensitive networks, it is thus expected that robustness to perturbation will be further strengthened if they are closer to normal, which is a phenomenon previously identified in fluid and drive-response systems~\cite{Trefethen1993,Illing2002}.

The experimental results presented here verify that in a network of diffusively-coupled oscillators the rate of convergence to
synchronization depends non-monotonically on the number of links. We also predict and experimentally demonstrate that,  depending on the eigenvector properties of the coupling matrix, co-spectral networks can exhibit quantitatively and qualitatively different convergence to synchrony. This study introduces the concept of sensitive and nonsensitive networks,  providing objective criteria for determining the robustness of real networks based on their eigenvector degeneracies.

We thank C.\  Williams for designing the amplifier circuit boards. This work was supported by DOD MURI Grant No.\ ONR N000140710734 and NSF Grant No.\ DMS-0709212.


\pagebreak

\begin{center}
{\bf Supplementary Information }
\end{center}
\renewcommand{\thetable}{{S\arabic{table}}}
\begin{table}[h]
\centering
\includegraphics{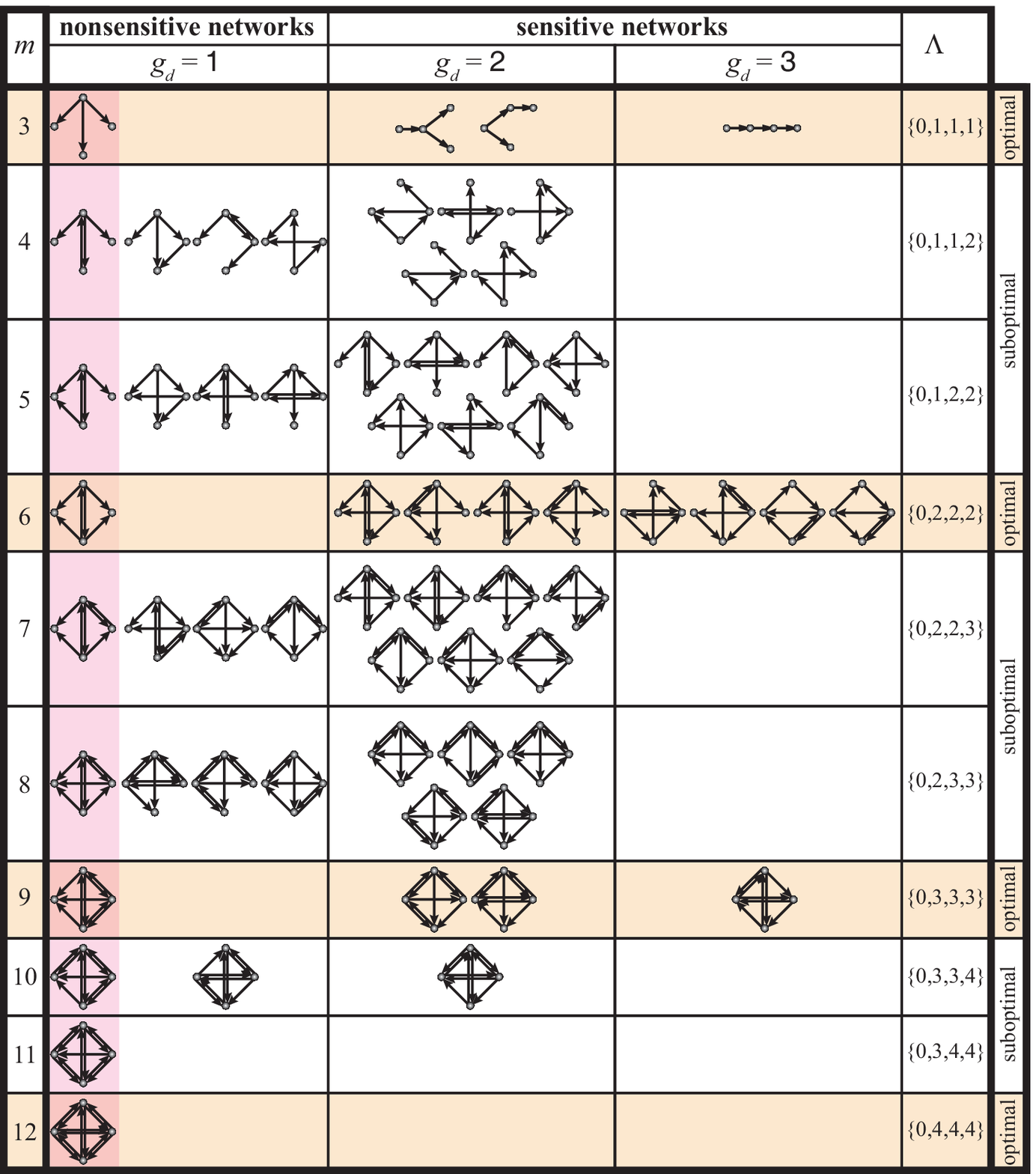}
\caption{Optimal (shaded rows) and suboptimal (white rows)  binary
networks with $N=4$ nodes classified according to the number of
connections $m$ (rows) and geometric degeneracy $g_d$ (columns).
The highlighted column (leftmost) shows a path (used in the
experiment of Fig.\ 2(c)) from an optimal tree ($m = 3$) to a
fully connected network ($m = 12$) which contains only
nonsensitive configurations ($g_d = 1$).} \label{suptable}
\end{table}

\renewcommand{\thefigure}{{S1}}
\begin{figure*}[h]
\centering
\includegraphics{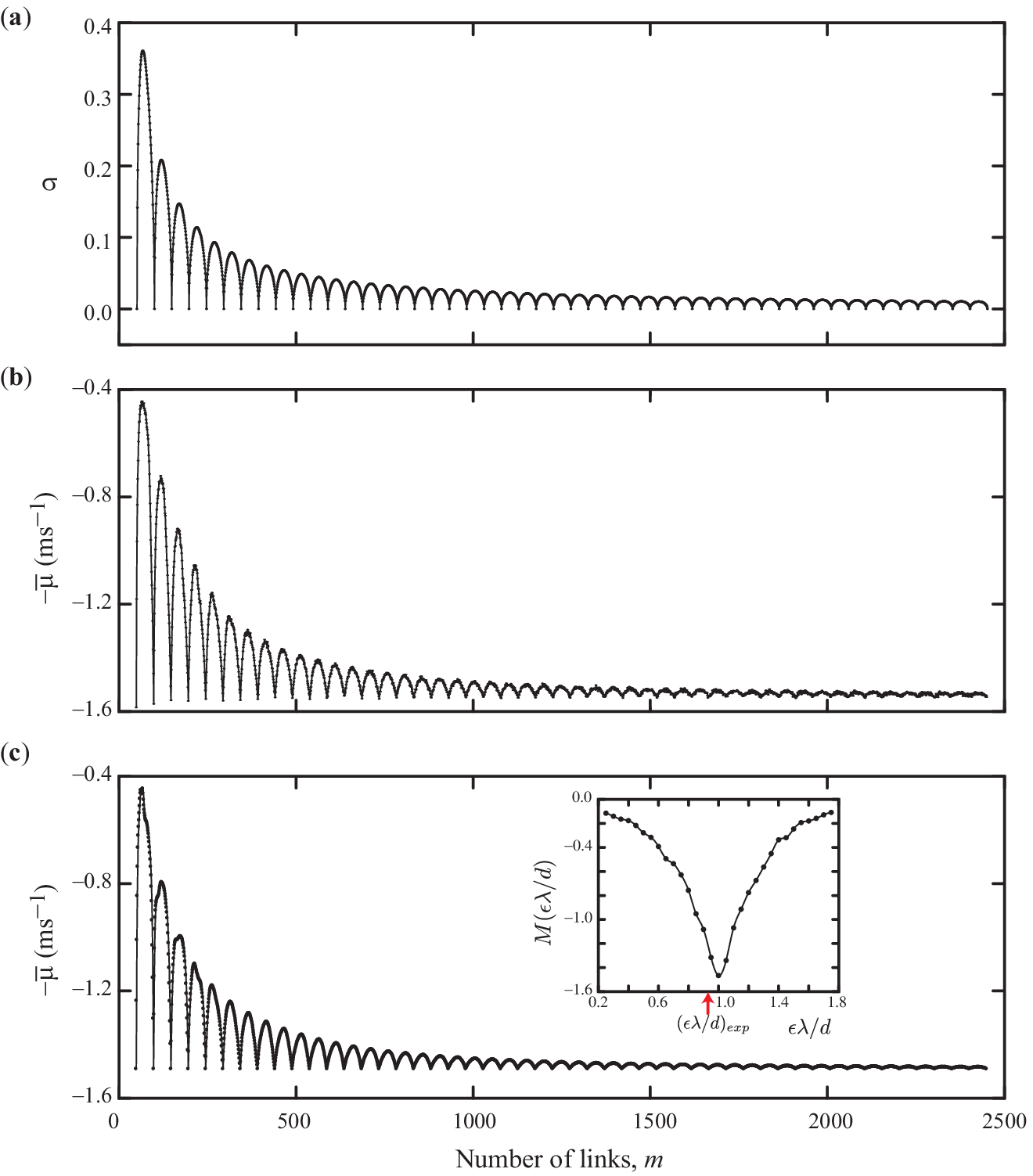}
\caption{Non-monotonic behavior of synchronization properties for
nonsensitive networks with $N=50$ nodes. (a) Eigenvalue spread
$\sigma$  as a function of the number of links $m$. (b) Mean
convergence rate  to synchronization, $\bar{\mu}$, determined by
simulating the system as modeled by Eqs.\ (1)-(2) (main text). (c)
Convergence rate estimated as a weighted average over the various
eigenmodes using the experimentally measured master stability
function $M(\epsilon\lambda/d)$ (inset), which was obtained from a
series of experimental measurements on a two node network. The
domain of $M$ shown corresponds to the region where
synchronization is observed experimentally. The coupling strength
$\epsilon$ used in (b, c) is chosen so that $\epsilon\lambda/d$
for the optimal networks of  size $50$ is the same as the
experimental value used for the optimal networks of size $4$. This
is marked with the subscript {\it exp} in the inset of panel (c).
} \label{supfig1}
\end{figure*}

\renewcommand{\thefigure}{{S2}}
\begin{figure*}[h]
\centering
\includegraphics[width=5in]{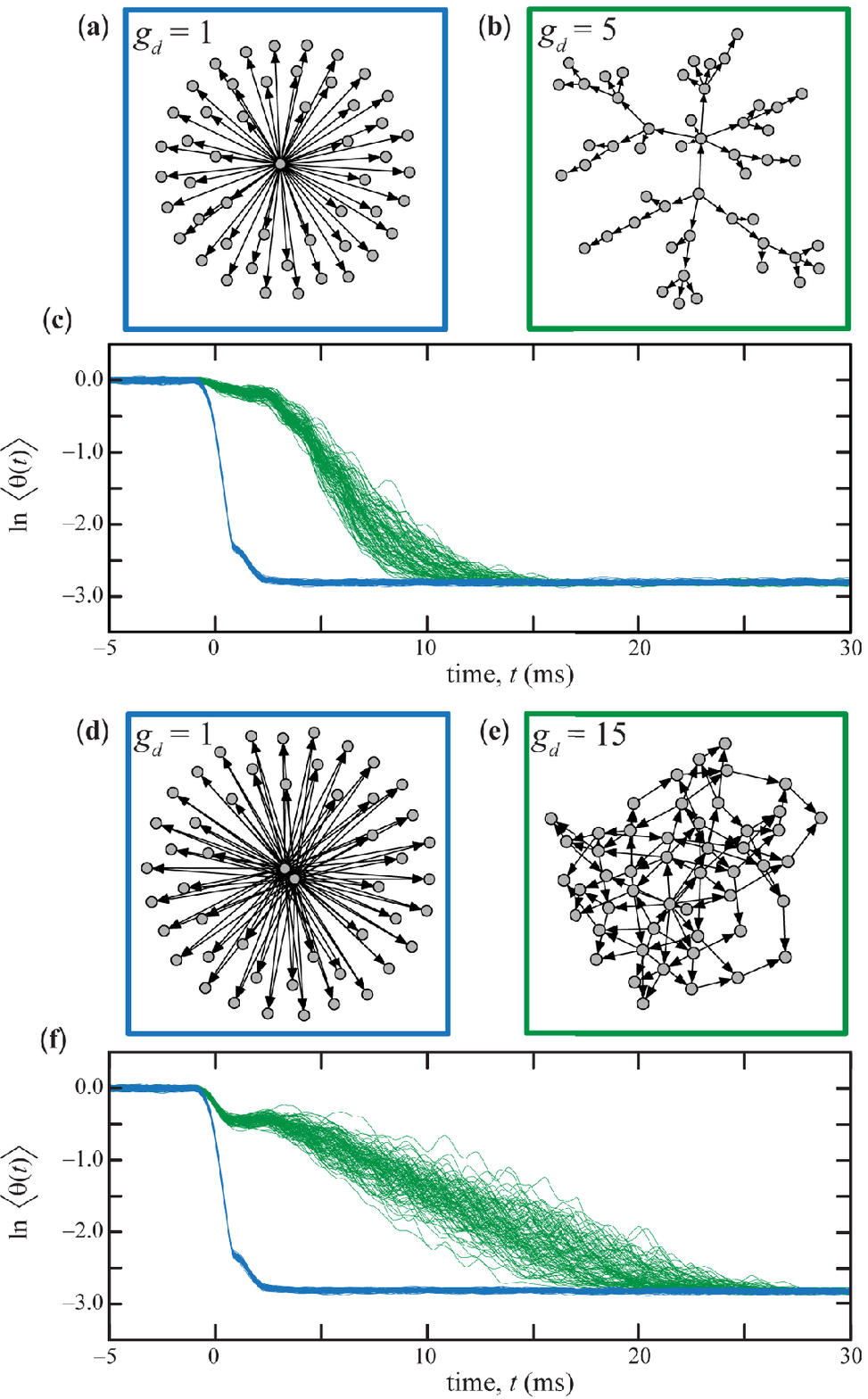}
\caption{Transition to synchronization for sensitive and
nonsensitive networks with $N=50$ nodes.  (a, b) Configurations
with $m=N-1$ links and $g_d=1$ and $5$, respectively.  (c)
Numerical simulation of the synchronization error,
$\langle\theta(t)\rangle$, for the networks shown in (a) ($g_d=1$,
blue) and (b) ($g_d=5$, green). (d-f) Same as in (a-c) for
configurations with $m=2N-2$ links,  and $g_d=1$ (blue) and $15$
(green).} \label{supfig2}
\end{figure*}

\renewcommand{\thefigure}{{S3}}
\begin{figure*}[h]
\centering
\includegraphics[width=6in]{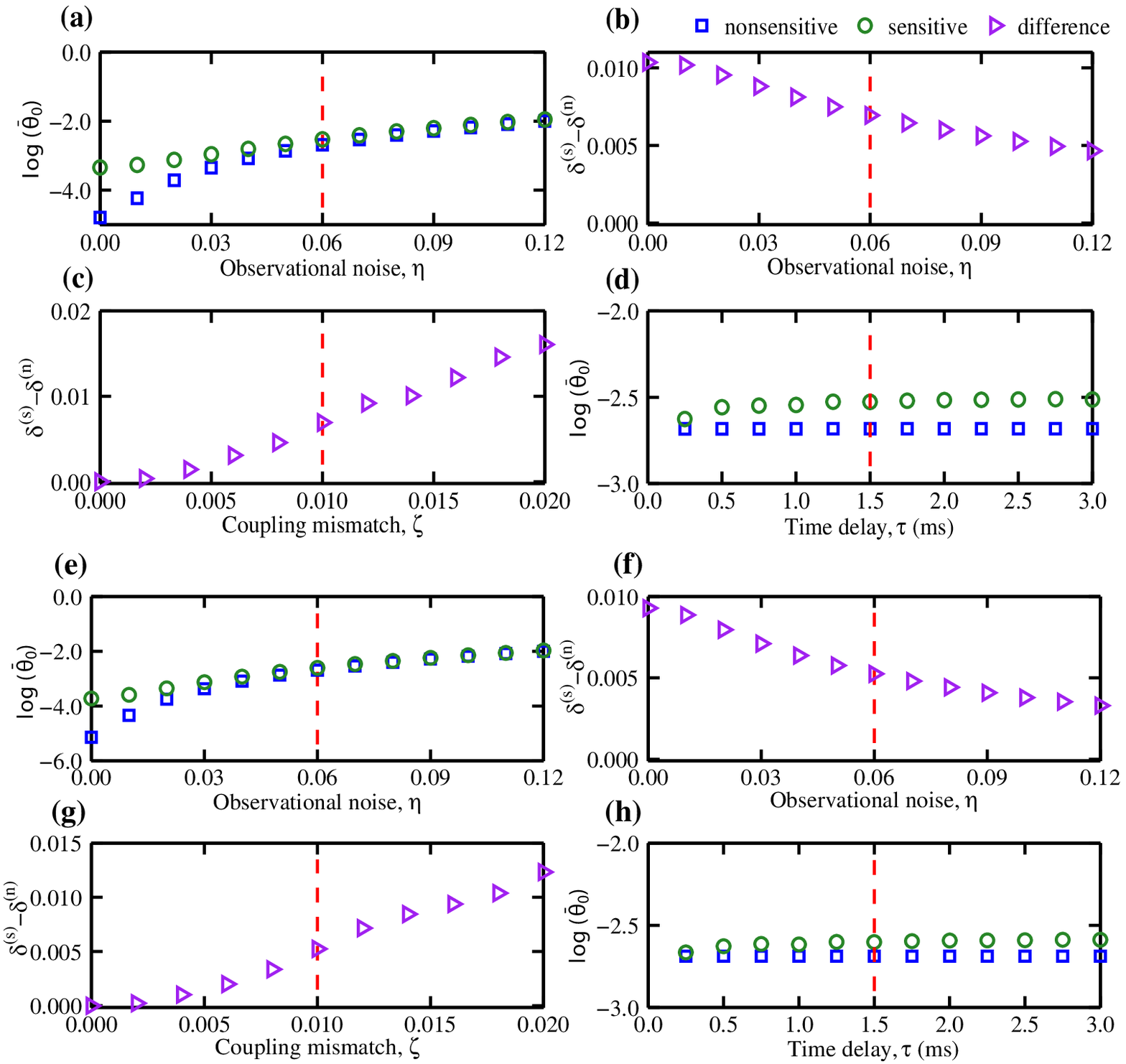}
\caption{Simulation results for networks with $N=50$ nodes. (a-d)
Counterpart of Fig.~4 for the networks considered in Fig.~S2(a-c).
(e-h) Corresponding results for the networks considered in
Fig.~S2(d-f). } \label{supfig3}
\end{figure*}

\end{document}